\documentclass[12pt]{article}

\usepackage[a4paper]{geometry}
\usepackage{amssymb}
\usepackage{latexsym}
\usepackage{graphicx}
\usepackage{epic}
\usepackage{eepic}
\usepackage[authoryear]{natbib}

\newlength{\myfigwidth}
\newcommand{\eqref}[1]{(\ref{#1})}

\begin{document}

\title{Firm Diversification and the Law of Proportionate
  Effect\thanks{Financial support from M.U.R.S.T. and from the Merck
    Foundation (E.P.R.I.S. project) is gratefully acknowledged}}
\author{Giulio Bottazzi\thanks{E-mail address: bottazzi@sssup.it }
  \\S.  Anna School of Advanced Studies, Pisa, Italy\\}

\maketitle

\begin{abstract}
  The paper presents an analysis performed over the worldwide top
  $150$ firms in the pharmaceutical industry. It begins with a test of
  the Gibrat's Law of Proportionate Effect finding, in line with
  previous literature, a violation concerning the variance of the
  growth. Then it shows, using disaggregated data on sub-markets
  (defined according to the ATC code) that this violation can be
  completely referred to the existence of a ``diversification''
  effect, namely a scale relation between the number of active
  sub-markets a firm posses and its size.  The observed scaling
  property of the firms diversification patterns is in contrast with
  the linear assumption typically made in literature.  Finally, to
  interpret the findings, the work proposes a stochastic model for
  firms diversification which fits quite well the data.
\end{abstract}

\section{Introduction}
\label{sec:intro}

In industrial economics the ``stochastic'' approach to the description
of firms behavior has been introduced long ago with the advent of the
Gibrat's ``random walk'' description of the business firm growth
process (\citet{Gibrat}; for an historical review see
\citet{sutton1}).  Although this kind of analysis has been applied to
the study of the time evolution of firms ``size'' (measured by diverse
variables like sales or employees) with a certain degree of success,
the persistent lack of data has prevented its application to other
firm-specific relevant economic variables (as productivity,
profitability or diversification, to cite a few).

Moreover, the assumption of ``identity'' between the ``objects''
(firms) that form the system (industry or sector) implicit in the
``statistical'' description, while perfectly natural in contexts like
biology, ecology or physics, where the ``stochastic'' models have
found a large domain of application, seemed to many authors
dissatisfying if applied to the description of economic phenomena. In
fact, a large mainstream literature seems going exactly in the
opposite direction, trying to characterize industry dynamics starting
from a set of firm-specific characteristics (both observable, as the
firm's age \citet{Mueller} or unobservable, as its ``state'' inside a
``natural'' growth cycle \citet{Greiner}). These models, contrary to
the ``stochastic'' ones, typically concentrate on variables, like
productivity or profitability, considered more relevant to describe
the actual firm ``performance''. It must be noted, however, that even
if these studies can be used to obtain ``interpretative guidelines''
(see for instance \cite{DMOS}) they would need, in order to be
(dis)proved or at least completely exploited in their descriptive
power, datasets far more complete that the ones available today.

The present contribution can be considered halfway between the two
previous approaches. Indeed, while it is based on ``stochastic
modeling'' which, with its high degree of ``essentiality'', has proved
to be a powerful tool for an empirical based analysis of firms size
dynamics, it extend such technique to an other, in same sense more
``strategic'', aspect of firms behavior constituted by their
diversification patterns. In this sense it constitutes a first attempt
to include, via an empirical grounded analysis, relevant firm-specific
economic variables in the description of firm growth performances.

The baseline model in stochastic firms growth dates back to the
pioneering work of Gibrat who proposed a model, known as
``Law of Proportionate Effect'' (LPE), relating the size of a firm and
its rate of growth by the expression
\begin{equation}
  \label{eq:gibrat}
  S(t+1)=S(t)\left(1+R(t)\right)
\end{equation}
where $S(t)$ is the firm's size at time $t$ and $R$ is a random
variable not dependent on $S$. This process reduces to a random walk
in the log of size $s(t)=\log(S(t))$ and (under the usual assumption
of validity of the Central Limit Theorem) predict an asymptotic log -
normal size distribution. This ``crude'' model misses many important
aspects of industrial dynamics (firms entry, exit, merging, etc.)
and various refinements have been proposed (as an example see
\citet{Hart}, \citet{SB}, \citet{IS}, \citet{sutton2} and recently
\citet{gerosky}) but it provides a robust framework for the
interpretation of data. From its introduction the LPE has been
empirically tested by many authors on different and heterogeneous
datasets (for a recent review see \citet{sutton1}) generally obtaining
two, in some sense opposite, conclusions: the Gibrat hypothesis is
confirmed in first approximation by the lack of any relationship
between the (log) size of firms and their rate of growth but is
violated by a clear dependence of the growth variance on size (see,
for instance \citet{evans,hall,mansfield} and more
recently~\citet{stanley1,stanley2}).

A natural explanation for the reduction of growth variance with size
could be a sort of ``portfolio'' effect (see eg. \citet{hympa}).  The
basic idea is that a firm can be described as a collection of
``atomic'' components (line of productions, plants, etc.) of roughly
the same size whose number would be proportional to the size of the
firm. Under the assumption that the growth processes for different
components are independent, the Law of Large Number (LLN) would
predict a linear relationship between the firm size and the variance
of its growth rate. The observed dependence is however milder and the
failure of the LLN is usually inputed (see e.g. \citet{boeri}) to the
existence of a ``relation'' between the firm components that makes the
aggregation of the ``atomic'' growths not simply additive\footnote{The
  sole introduction of correlation in growth components is clearly not
  enough.} A model has recently been proposed
\citet{stanley1,stanley2}, based on a supposed intra-firm complex
hierarchical structure, which, opportunely tuned, reproduce quite well
the observed behavior. On the contrary, I will show in what follows
that, if one correctly identifies the ``atomic'' contributions to
growth and the scaling relation between their number and the firm
size, the LLN does a good job in explaining the observed Gibrat
violation, at least as far as the database under analysis is
concerned, without any need of intra-firm ``structure''.

The analysis presented in the following is based on the dataset PHID
(Pharmaceutical Industry Database)\footnote{Developed at the
  University of Siena} that covers the top $100$ firms relative to the
seven major western markets (USA, United Kingdom, France, Germany,
Spain, Italy and Canada) in the period ranging from $1987$ to $1997$.
The dataset provides the firms sales in USD, disaggregated up to the
4-digit-level of the Anatomical Therapeutic Classification scheme
(ATC) in 517 micro classes.

The firms under analysis are obtained aggregating the regional figures
and
merging from the beginning the sales pertaining to firms that belong
to the same entity at the end of period. Moreover, I consider only the
$150$ so obtained firms with the top sales at the beginning of the
period\footnote{This cut is imposed to minimize the possibility of
  selection bias due to the different ``relative'' sizes of national
  markets. It does however introduce the interpretative problem of
  studying the evolution of a subset of firms that, at later time, are
  in general sparse in the ranked set.  I have checked that this
  effect is almost negligible, due to the low rank mobility of the
  industry, yet it is present and will shape the lower tail of the
  size distribution.}.

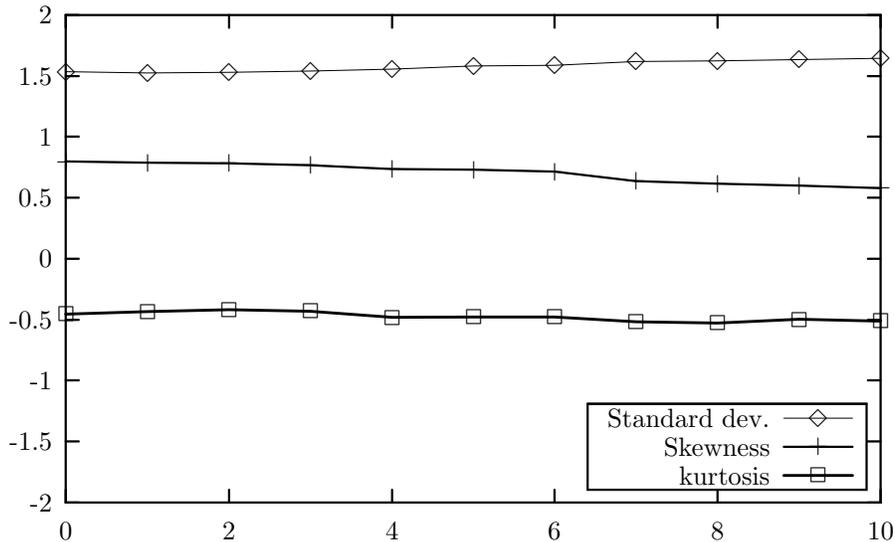
\begin{figure}[tbp]
  \begin{center}
\setlength{\unitlength}{0.240900pt}
\begin{picture}(1500,900)(0,0)
\footnotesize
\thicklines \path(154,90)(174,90)
\thicklines \path(1433,90)(1413,90)
\put(132,90){\makebox(0,0)[r]{-2}}
\thicklines \path(154,186)(174,186)
\thicklines \path(1433,186)(1413,186)
\put(132,186){\makebox(0,0)[r]{-1.5}}
\thicklines \path(154,282)(174,282)
\thicklines \path(1433,282)(1413,282)
\put(132,282){\makebox(0,0)[r]{-1}}
\thicklines \path(154,377)(174,377)
\thicklines \path(1433,377)(1413,377)
\put(132,377){\makebox(0,0)[r]{-0.5}}
\thicklines \path(154,473)(174,473)
\thicklines \path(1433,473)(1413,473)
\put(132,473){\makebox(0,0)[r]{0}}
\thicklines \path(154,569)(174,569)
\thicklines \path(1433,569)(1413,569)
\put(132,569){\makebox(0,0)[r]{0.5}}
\thicklines \path(154,665)(174,665)
\thicklines \path(1433,665)(1413,665)
\put(132,665){\makebox(0,0)[r]{1}}
\thicklines \path(154,760)(174,760)
\thicklines \path(1433,760)(1413,760)
\put(132,760){\makebox(0,0)[r]{1.5}}
\thicklines \path(154,856)(174,856)
\thicklines \path(1433,856)(1413,856)
\put(132,856){\makebox(0,0)[r]{2}}
\thicklines \path(154,90)(154,110)
\thicklines \path(154,856)(154,836)
\put(154,45){\makebox(0,0){0}}
\thicklines \path(410,90)(410,110)
\thicklines \path(410,856)(410,836)
\put(410,45){\makebox(0,0){2}}
\thicklines \path(666,90)(666,110)
\thicklines \path(666,856)(666,836)
\put(666,45){\makebox(0,0){4}}
\thicklines \path(921,90)(921,110)
\thicklines \path(921,856)(921,836)
\put(921,45){\makebox(0,0){6}}
\thicklines \path(1177,90)(1177,110)
\thicklines \path(1177,856)(1177,836)
\put(1177,45){\makebox(0,0){8}}
\thicklines \path(1433,90)(1433,110)
\thicklines \path(1433,856)(1433,836)
\put(1433,45){\makebox(0,0){10}}
\thicklines \path(154,90)(1433,90)(1433,856)(154,856)(154,90)
\thicklines \path(973,110)(973,245)(1411,245)(1411,110)(973,110)
\thicklines \path(973,245)(1411,245)
\put(1259,223){\makebox(0,0)[r]{Standard dev.}}
\thinlines \path(1281,223)(1389,223)
\thinlines \path(154,767)(154,767)(282,765)(410,766)(538,768)(666,771)(794,776)(921,777)(1049,783)(1177,784)(1305,786)(1433,788)
\put(154,767){\raisebox{-1.2pt}{\makebox(0,0){$\Diamond$}}}
\put(282,765){\raisebox{-1.2pt}{\makebox(0,0){$\Diamond$}}}
\put(410,766){\raisebox{-1.2pt}{\makebox(0,0){$\Diamond$}}}
\put(538,768){\raisebox{-1.2pt}{\makebox(0,0){$\Diamond$}}}
\put(666,771){\raisebox{-1.2pt}{\makebox(0,0){$\Diamond$}}}
\put(794,776){\raisebox{-1.2pt}{\makebox(0,0){$\Diamond$}}}
\put(921,777){\raisebox{-1.2pt}{\makebox(0,0){$\Diamond$}}}
\put(1049,783){\raisebox{-1.2pt}{\makebox(0,0){$\Diamond$}}}
\put(1177,784){\raisebox{-1.2pt}{\makebox(0,0){$\Diamond$}}}
\put(1305,786){\raisebox{-1.2pt}{\makebox(0,0){$\Diamond$}}}
\put(1433,788){\raisebox{-1.2pt}{\makebox(0,0){$\Diamond$}}}
\put(1335,223){\raisebox{-1.2pt}{\makebox(0,0){$\Diamond$}}}
\put(1259,178){\makebox(0,0)[r]{Skewness}}
\thicklines \path(1281,178)(1389,178)
\thicklines \path(154,626)(154,626)(282,624)(410,623)(538,620)(666,614)(794,613)(921,610)(1049,595)(1177,591)(1305,588)(1433,584)
\put(154,626){\makebox(0,0){$+$}}
\put(282,624){\makebox(0,0){$+$}}
\put(410,623){\makebox(0,0){$+$}}
\put(538,620){\makebox(0,0){$+$}}
\put(666,614){\makebox(0,0){$+$}}
\put(794,613){\makebox(0,0){$+$}}
\put(921,610){\makebox(0,0){$+$}}
\put(1049,595){\makebox(0,0){$+$}}
\put(1177,591){\makebox(0,0){$+$}}
\put(1305,588){\makebox(0,0){$+$}}
\put(1433,584){\makebox(0,0){$+$}}
\put(1335,178){\makebox(0,0){$+$}}
\put(1259,133){\makebox(0,0)[r]{kurtosis}}
\Thicklines \path(1281,133)(1389,133)
\Thicklines \path(154,386)(154,386)(282,390)(410,393)(538,391)(666,381)(794,382)(921,382)(1049,374)(1177,372)(1305,378)(1433,375)
\put(154,386){\raisebox{-1.2pt}{\makebox(0,0){$\Box$}}}
\put(282,390){\raisebox{-1.2pt}{\makebox(0,0){$\Box$}}}
\put(410,393){\raisebox{-1.2pt}{\makebox(0,0){$\Box$}}}
\put(538,391){\raisebox{-1.2pt}{\makebox(0,0){$\Box$}}}
\put(666,381){\raisebox{-1.2pt}{\makebox(0,0){$\Box$}}}
\put(794,382){\raisebox{-1.2pt}{\makebox(0,0){$\Box$}}}
\put(921,382){\raisebox{-1.2pt}{\makebox(0,0){$\Box$}}}
\put(1049,374){\raisebox{-1.2pt}{\makebox(0,0){$\Box$}}}
\put(1177,372){\raisebox{-1.2pt}{\makebox(0,0){$\Box$}}}
\put(1305,378){\raisebox{-1.2pt}{\makebox(0,0){$\Box$}}}
\put(1433,375){\raisebox{-1.2pt}{\makebox(0,0){$\Box$}}}
\put(1335,133){\raisebox{-1.2pt}{\makebox(0,0){$\Box$}}}
\end{picture}
    \caption{Moments of the $g$ distribution computed at different times}
    \label{fig:g_moments}
  \end{center}
\end{figure}

In the next section I present a brief overview of the statistical
properties of the size distribution and of the growth process.  A
wider analysis is presented in \citet{BDLPR}, here I concentrate the
attention on the violation of the Gibrat model emerging as a relation
between the variance of growth and the size of a firm. In
Section~\ref{sec:gibrat}, I perform a deeper analysis of this
relation, using disaggregate data on the ``sub-markets'' defined by
the 4-digit ATC code. I am then able to reduce the Gibrat violation to
a ``diversification'' effect, due to a scale relationship between the
firm size and the number of its active markets. Finally in
Section~\ref{sec:branching}, I propose a model, inspired by the
structure of the ``technological evolution'' in the pharmaceutical
industry, which accounts for the observed pattern of firm's
diversification.

\section{Data analysis}
\label{sec:data}

\begin{figure}[tbp]
  \begin{center}
\setlength{\unitlength}{0.240900pt}
\begin{picture}(1500,900)(0,0)
\footnotesize
\thicklines \path(132,90)(152,90)
\thicklines \path(1433,90)(1413,90)
\put(110,90){\makebox(0,0)[r]{0}}
\thicklines \path(132,167)(152,167)
\thicklines \path(1433,167)(1413,167)
\put(110,167){\makebox(0,0)[r]{0.1}}
\thicklines \path(132,243)(152,243)
\thicklines \path(1433,243)(1413,243)
\put(110,243){\makebox(0,0)[r]{0.2}}
\thicklines \path(132,320)(152,320)
\thicklines \path(1433,320)(1413,320)
\put(110,320){\makebox(0,0)[r]{0.3}}
\thicklines \path(132,396)(152,396)
\thicklines \path(1433,396)(1413,396)
\put(110,396){\makebox(0,0)[r]{0.4}}
\thicklines \path(132,473)(152,473)
\thicklines \path(1433,473)(1413,473)
\put(110,473){\makebox(0,0)[r]{0.5}}
\thicklines \path(132,550)(152,550)
\thicklines \path(1433,550)(1413,550)
\put(110,550){\makebox(0,0)[r]{0.6}}
\thicklines \path(132,626)(152,626)
\thicklines \path(1433,626)(1413,626)
\put(110,626){\makebox(0,0)[r]{0.7}}
\thicklines \path(132,703)(152,703)
\thicklines \path(1433,703)(1413,703)
\put(110,703){\makebox(0,0)[r]{0.8}}
\thicklines \path(132,779)(152,779)
\thicklines \path(1433,779)(1413,779)
\put(110,779){\makebox(0,0)[r]{0.9}}
\thicklines \path(132,856)(152,856)
\thicklines \path(1433,856)(1413,856)
\put(110,856){\makebox(0,0)[r]{1}}
\thicklines \path(132,90)(132,110)
\thicklines \path(132,856)(132,836)
\put(132,45){\makebox(0,0){-5}}
\thicklines \path(295,90)(295,110)
\thicklines \path(295,856)(295,836)
\put(295,45){\makebox(0,0){-4}}
\thicklines \path(457,90)(457,110)
\thicklines \path(457,856)(457,836)
\put(457,45){\makebox(0,0){-3}}
\thicklines \path(620,90)(620,110)
\thicklines \path(620,856)(620,836)
\put(620,45){\makebox(0,0){-2}}
\thicklines \path(783,90)(783,110)
\thicklines \path(783,856)(783,836)
\put(783,45){\makebox(0,0){-1}}
\thicklines \path(945,90)(945,110)
\thicklines \path(945,856)(945,836)
\put(945,45){\makebox(0,0){0}}
\thicklines \path(1108,90)(1108,110)
\thicklines \path(1108,856)(1108,836)
\put(1108,45){\makebox(0,0){1}}
\thicklines \path(1270,90)(1270,110)
\thicklines \path(1270,856)(1270,836)
\put(1270,45){\makebox(0,0){2}}
\thicklines \path(1433,90)(1433,110)
\thicklines \path(1433,856)(1433,836)
\put(1433,45){\makebox(0,0){3}}
\thicklines \path(132,90)(1433,90)(1433,856)(132,856)(132,90)
\thicklines \path(1105,110)(1105,245)(1411,245)(1411,110)(1105,110)
\thicklines \path(1105,245)(1411,245)
\put(1259,223){\makebox(0,0)[r]{F(g,1)}}
\thinlines \path(1281,223)(1389,223)
\thinlines \path(412,95)(412,95)(412,100)(422,105)(427,110)(434,115)(438,120)(438,126)(444,131)(452,136)(453,141)(456,146)(459,151)(461,156)(461,161)(462,166)(462,171)(468,176)(470,181)(471,186)(477,191)(478,197)(485,202)(485,207)(488,212)(489,217)(491,222)(492,227)(496,232)(496,237)(497,242)(498,247)(503,252)(507,257)(512,262)(512,268)(518,273)(525,278)(525,283)(526,288)(530,293)(534,298)(534,303)(535,308)(539,313)(540,318)(540,323)(543,328)(544,333)(545,339)(552,344)
\thinlines \path(552,344)(555,349)(560,354)(560,359)(560,364)(563,369)(569,374)(572,379)(572,384)(577,389)(583,394)(584,399)(587,405)(596,410)(601,415)(613,420)(615,425)(616,430)(618,435)(647,440)(652,445)(653,450)(657,455)(660,460)(660,465)(663,470)(669,476)(671,481)(674,486)(681,491)(682,496)(686,501)(695,506)(697,511)(700,516)(707,521)(709,526)(710,531)(720,536)(735,541)(739,547)(741,552)(741,557)(744,562)(763,567)(764,572)(766,577)(770,582)(776,587)(782,592)(793,597)
\thinlines \path(793,597)(811,602)(812,607)(824,613)(832,618)(834,623)(840,628)(841,633)(858,638)(858,643)(869,648)(871,653)(872,658)(872,663)(876,668)(878,673)(884,678)(901,684)(902,689)(926,694)(931,699)(935,704)(947,709)(953,714)(960,719)(984,724)(1000,729)(1012,734)(1054,739)(1066,744)(1106,749)(1108,755)(1123,760)(1126,765)(1133,770)(1136,775)(1162,780)(1165,785)(1190,790)(1191,795)(1191,800)(1214,805)(1223,810)(1251,815)(1251,820)(1259,826)(1276,831)(1278,836)(1284,841)(1307,846)(1318,851)
\put(1259,178){\makebox(0,0)[r]{F(g,5)}}
\thicklines \path(1281,178)(1389,178)
\thicklines \path(341,95)(341,95)(364,100)(381,105)(395,110)(397,115)(408,120)(418,126)(421,131)(427,136)(437,141)(439,146)(443,151)(449,156)(450,161)(462,166)(467,171)(467,176)(470,181)(470,186)(471,191)(472,197)(476,202)(480,207)(483,212)(484,217)(484,222)(490,227)(493,232)(495,237)(497,242)(500,247)(506,252)(506,257)(509,262)(509,268)(512,273)(517,278)(521,283)(526,288)(526,293)(527,298)(527,303)(529,308)(532,313)(533,318)(535,323)(536,328)(542,333)(548,339)(561,344)
\thicklines \path(561,344)(564,349)(564,354)(566,359)(575,364)(577,369)(582,374)(584,379)(586,384)(590,389)(590,394)(603,399)(604,405)(605,410)(618,415)(620,420)(635,425)(644,430)(649,435)(654,440)(657,445)(664,450)(664,455)(667,460)(681,465)(684,470)(690,476)(691,481)(691,486)(693,491)(693,496)(695,501)(696,506)(697,511)(700,516)(708,521)(710,526)(714,531)(714,536)(718,541)(727,547)(730,552)(737,557)(742,562)(750,567)(755,572)(763,577)(768,582)(770,587)(774,592)(791,597)
\thicklines \path(791,597)(794,602)(795,607)(802,613)(809,618)(823,623)(830,628)(842,633)(842,638)(842,643)(844,648)(849,653)(873,658)(877,663)(879,668)(884,673)(887,678)(892,684)(903,689)(923,694)(947,699)(953,704)(990,709)(994,714)(995,719)(1000,724)(1019,729)(1021,734)(1051,739)(1055,744)(1093,749)(1101,755)(1108,760)(1131,765)(1131,770)(1145,775)(1151,780)(1178,785)(1185,790)(1188,795)(1204,800)(1210,805)(1213,810)(1227,815)(1262,820)(1272,826)(1273,831)(1274,836)(1277,841)(1303,846)(1324,851)
\put(1259,133){\makebox(0,0)[r]{F(g,10)}}
\Thicklines \path(1281,133)(1389,133)
\Thicklines \path(219,95)(219,95)(292,100)(319,105)(325,110)(333,115)(335,120)(339,126)(350,131)(356,136)(381,141)(390,146)(392,151)(392,156)(399,161)(399,166)(402,171)(407,176)(411,181)(425,186)(438,191)(448,197)(455,202)(457,207)(458,212)(459,217)(467,222)(467,227)(468,232)(478,237)(479,242)(480,247)(481,252)(483,257)(487,262)(489,268)(492,273)(507,278)(509,283)(524,288)(530,293)(531,298)(537,303)(540,308)(542,313)(550,318)(550,323)(552,328)(555,333)(562,339)(564,344)
\Thicklines \path(564,344)(565,349)(568,354)(571,359)(574,364)(578,369)(580,374)(584,379)(587,384)(590,389)(599,394)(601,399)(607,405)(609,410)(610,415)(612,420)(619,425)(622,430)(633,435)(633,440)(636,445)(637,450)(638,455)(644,460)(647,465)(648,470)(652,476)(654,481)(663,486)(665,491)(670,496)(674,501)(685,506)(686,511)(695,516)(700,521)(701,526)(709,531)(718,536)(722,541)(724,547)(739,552)(743,557)(750,562)(775,567)(788,572)(791,577)(799,582)(800,587)(805,592)(807,597)
\Thicklines \path(807,597)(808,602)(812,607)(812,613)(819,618)(823,623)(823,628)(827,633)(832,638)(845,643)(852,648)(853,653)(855,658)(864,663)(867,668)(884,673)(897,678)(910,684)(910,689)(916,694)(926,699)(938,704)(953,709)(956,714)(956,719)(972,724)(978,729)(990,734)(1013,739)(1066,744)(1074,749)(1102,755)(1103,760)(1119,765)(1137,770)(1155,775)(1191,780)(1200,785)(1202,790)(1203,795)(1213,800)(1215,805)(1224,810)(1242,815)(1260,820)(1267,826)(1270,831)(1276,836)(1279,841)(1286,846)(1324,851)
\end{picture}
    \caption{Distribution function for the firm ``normalized size'' $g$.
      The three curves correspond to the distribution obtained for
      $t=0,5,10$.}
    \label{fig:g_distrib}
  \end{center}
\end{figure}
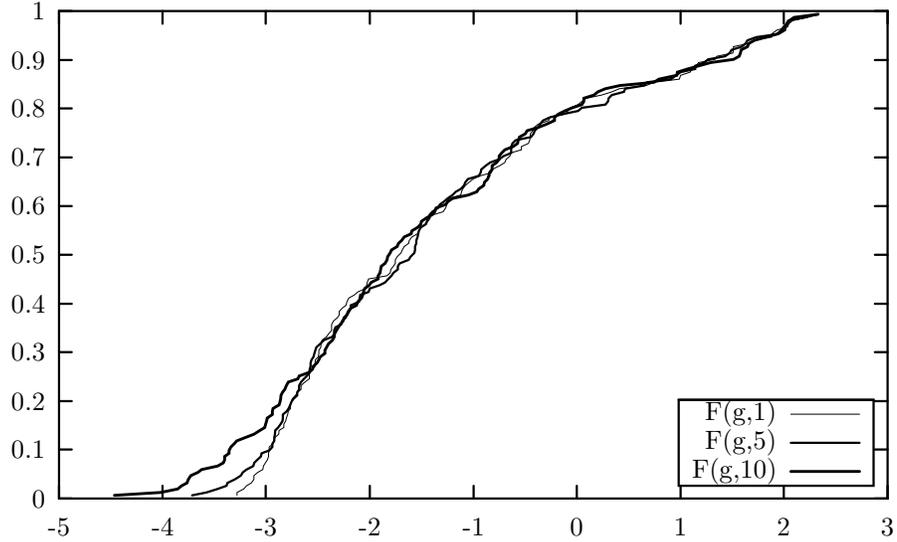

Let $S_i(t)$ be the size of firm $i$ ($i\in[1,\ldots,N]$) at time $t$
($i\in[0,\ldots,T]$) where (see Sec.~\ref{sec:intro}) $N=150$ and $T=10$
and let define the `normalized sizes'' as
$G_i(t)=S_i(t)/<S_i(t)>$. This variable and its log
$g_i(t)=\log(G_i(t))$ are characterized by distribution functions that
can be considered stationary in time.  The reliability of this
assumption can be checked plotting the moments of the distribution as
a function of time (see Fig.~\ref{fig:g_moments}): a part from a
constant mild increase in the variance of the distribution, the
approximation turns out to be good.  However the major contribution to
the variance increase comes clearly from the low $g$ region ( see
Fig.~\ref{fig:g_distrib}) where a ``spurious'' (due to the selection
bias) diffusion toward smaller sizes is present. This diffusion is
responsible for the observed increase in variance. Once understood,
this effect is statistically irrelevant and in the following analysis
the $g$ variable is thought identically distributed at every time
step.\footnote{By the way, note that the transition from the $S$ to
  the $G$ variables, makes the discussion insensitive to any prices
  variation that is constant over all the pharmaceutical products (for
  instance, the monetary inflation or any price dynamics relatives to
  pharmaceutical industry).}

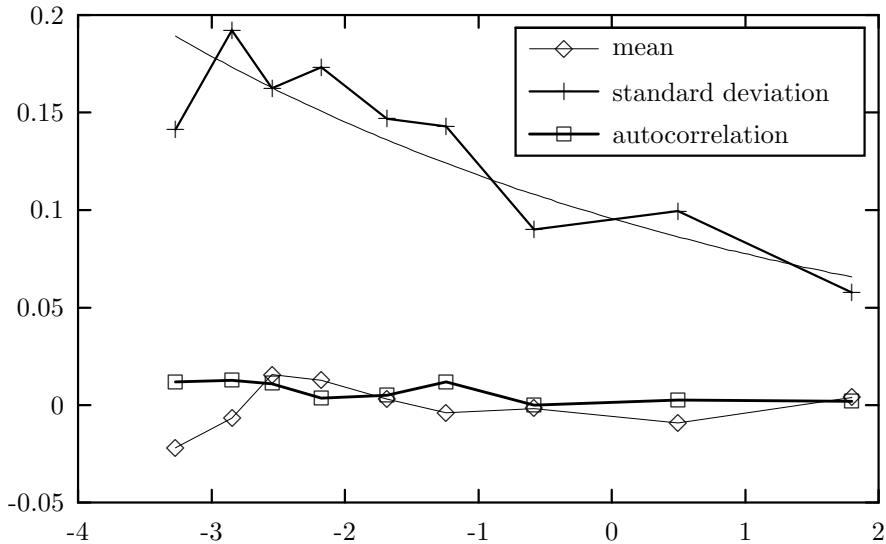
\begin{figure}[tbp]
  \begin{center}
\setlength{\unitlength}{0.240900pt}
\begin{picture}(1500,900)(0,0)
\footnotesize
\thicklines \path(176,90)(196,90)
\thicklines \path(1433,90)(1413,90)
\put(154,90){\makebox(0,0)[r]{-0.05}}
\thicklines \path(176,243)(196,243)
\thicklines \path(1433,243)(1413,243)
\put(154,243){\makebox(0,0)[r]{0}}
\thicklines \path(176,396)(196,396)
\thicklines \path(1433,396)(1413,396)
\put(154,396){\makebox(0,0)[r]{0.05}}
\thicklines \path(176,550)(196,550)
\thicklines \path(1433,550)(1413,550)
\put(154,550){\makebox(0,0)[r]{0.1}}
\thicklines \path(176,703)(196,703)
\thicklines \path(1433,703)(1413,703)
\put(154,703){\makebox(0,0)[r]{0.15}}
\thicklines \path(176,856)(196,856)
\thicklines \path(1433,856)(1413,856)
\put(154,856){\makebox(0,0)[r]{0.2}}
\thicklines \path(176,90)(176,110)
\thicklines \path(176,856)(176,836)
\put(176,45){\makebox(0,0){-4}}
\thicklines \path(386,90)(386,110)
\thicklines \path(386,856)(386,836)
\put(386,45){\makebox(0,0){-3}}
\thicklines \path(595,90)(595,110)
\thicklines \path(595,856)(595,836)
\put(595,45){\makebox(0,0){-2}}
\thicklines \path(805,90)(805,110)
\thicklines \path(805,856)(805,836)
\put(805,45){\makebox(0,0){-1}}
\thicklines \path(1014,90)(1014,110)
\thicklines \path(1014,856)(1014,836)
\put(1014,45){\makebox(0,0){0}}
\thicklines \path(1224,90)(1224,110)
\thicklines \path(1224,856)(1224,836)
\put(1224,45){\makebox(0,0){1}}
\thicklines \path(1433,90)(1433,110)
\thicklines \path(1433,856)(1433,836)
\put(1433,45){\makebox(0,0){2}}
\thicklines \path(176,90)(1433,90)(1433,856)(176,856)(176,90)
\thicklines \path(863,635)(863,836)(1411,836)(1411,635)(863,635)
\thicklines \path(863,836)(1411,836)
\put(1015,803){\makebox(0,0)[l]{mean}}
\thinlines \path(885,803)(993,803)
\thinlines \path(1391,255)(1391,255)(1118,215)(892,238)(754,231)(661,253)(558,282)(481,291)(418,223)(329,176)
\put(1391,255){\raisebox{-1.2pt}{\makebox(0,0){$\Diamond$}}}
\put(1118,215){\raisebox{-1.2pt}{\makebox(0,0){$\Diamond$}}}
\put(892,238){\raisebox{-1.2pt}{\makebox(0,0){$\Diamond$}}}
\put(754,231){\raisebox{-1.2pt}{\makebox(0,0){$\Diamond$}}}
\put(661,253){\raisebox{-1.2pt}{\makebox(0,0){$\Diamond$}}}
\put(558,282){\raisebox{-1.2pt}{\makebox(0,0){$\Diamond$}}}
\put(481,291){\raisebox{-1.2pt}{\makebox(0,0){$\Diamond$}}}
\put(418,223){\raisebox{-1.2pt}{\makebox(0,0){$\Diamond$}}}
\put(329,176){\raisebox{-1.2pt}{\makebox(0,0){$\Diamond$}}}
\put(939,803){\raisebox{-1.2pt}{\makebox(0,0){$\Diamond$}}}
\put(1015,736){\makebox(0,0)[l]{standard deviation}}
\thicklines \path(885,736)(993,736)
\thicklines \path(1391,420)(1391,420)(1118,548)(892,519)(754,681)(661,693)(558,774)(481,741)(418,832)(329,676)
\put(1391,420){\makebox(0,0){$+$}}
\put(1118,548){\makebox(0,0){$+$}}
\put(892,519){\makebox(0,0){$+$}}
\put(754,681){\makebox(0,0){$+$}}
\put(661,693){\makebox(0,0){$+$}}
\put(558,774){\makebox(0,0){$+$}}
\put(481,741){\makebox(0,0){$+$}}
\put(418,832){\makebox(0,0){$+$}}
\put(329,676){\makebox(0,0){$+$}}
\put(939,736){\makebox(0,0){$+$}}
\put(1015,669){\makebox(0,0)[l]{autocorrelation}}
\Thicklines \path(885,669)(993,669)
\Thicklines \path(1391,249)(1391,249)(1118,251)(892,243)(754,280)(661,259)(558,254)(481,277)(418,282)(329,280)
\put(1391,249){\raisebox{-1.2pt}{\makebox(0,0){$\Box$}}}
\put(1118,251){\raisebox{-1.2pt}{\makebox(0,0){$\Box$}}}
\put(892,243){\raisebox{-1.2pt}{\makebox(0,0){$\Box$}}}
\put(754,280){\raisebox{-1.2pt}{\makebox(0,0){$\Box$}}}
\put(661,259){\raisebox{-1.2pt}{\makebox(0,0){$\Box$}}}
\put(558,254){\raisebox{-1.2pt}{\makebox(0,0){$\Box$}}}
\put(481,277){\raisebox{-1.2pt}{\makebox(0,0){$\Box$}}}
\put(418,282){\raisebox{-1.2pt}{\makebox(0,0){$\Box$}}}
\put(329,280){\raisebox{-1.2pt}{\makebox(0,0){$\Box$}}}
\put(939,669){\raisebox{-1.2pt}{\makebox(0,0){$\Box$}}}
\thinlines \path(329,823)(329,823)(339,817)(350,811)(361,805)(371,799)(382,793)(393,787)(404,782)(414,776)(425,770)(436,765)(447,759)(457,754)(468,748)(479,743)(489,737)(500,732)(511,727)(522,722)(532,717)(543,712)(554,707)(565,702)(575,697)(586,692)(597,687)(607,683)(618,678)(629,673)(640,669)(650,664)(661,660)(672,655)(683,651)(693,647)(704,642)(715,638)(725,634)(736,630)(747,626)(758,622)(768,618)(779,614)(790,610)(801,606)(811,602)(822,598)(833,594)(843,591)(854,587)
\thinlines \path(854,587)(865,583)(876,580)(886,576)(897,573)(908,569)(919,566)(929,562)(940,559)(951,555)(961,552)(972,549)(983,546)(994,542)(1004,539)(1015,536)(1026,533)(1037,530)(1047,527)(1058,524)(1069,521)(1079,518)(1090,515)(1101,512)(1112,509)(1122,506)(1133,504)(1144,501)(1155,498)(1165,495)(1176,493)(1187,490)(1197,487)(1208,485)(1219,482)(1230,480)(1240,477)(1251,475)(1262,472)(1273,470)(1283,467)(1294,465)(1305,463)(1315,460)(1326,458)(1337,456)(1348,454)(1358,451)(1369,449)(1380,447)(1391,445)
\end{picture}
    \caption{Mean, standard deviation and autocorrelation
      of growth $h$ computed for different size bins plotted against
      the average size in the bin. The exponential fit to the standard
      deviation gives a value $\protect\gamma =-0.20+/-0.03$}
    \label{fig:bysize}
  \end{center}
\end{figure}

As a first step in the search for a violation of the LPE it is natural
to investigate the possible presence of a relation between the first
moments of the growth distribution and the size. One can proceed in a
straightforward way: said $h_i(t)=g_i(t+1)-g_i(t)$ the growth of firm
$i$ at time $t$, one bins all the firm according to their size
$g_i(t)$ in equally numbered sets, and computes the mean, standard
deviation and $1$-lag autocorrelation of the growth $h_i(t)$ for each
set separately. The results are reported in Fig.~\ref{fig:bysize}
against the average size of each bin.

Both the mean growth and the autocorrelation do not show any
particular pattern but
a clear dependence of the growth variance on size emerges and The Law
of Proportionate Effect is violated.  Fitting the relation between
variance of growth and size with an exponential law
\begin{equation}
  \label{eq:varsize}
  \sigma(h) \sim e^{\beta g}
\end{equation}
one find a value $\beta=.2 +- .03$ that is striking similar to the one
found in other analysis on different datasets
\citet{stanley1}~\citet{stanley2} and much lower then the ``portfolio''
prediction $\beta=.5$ discussed in Sec.~\ref{sec:intro}.

In what follows I will show that the LLN is, however, sufficient to
completely characterize the relation in \eqref{eq:varsize}. The key
point will be to look at the whole firm growth as an ``aggregation''
of its growth in the different sub-markets in which it operates.

\section{Diversification as a source of Gibrat violation}
\label{sec:gibrat}

One of the peculiar feature of the PHID database is the possibility of
disaggregate the sales figures until the $4$th digit of the ATC
code. This level of disaggregation allows to identify sub-markets that
are ``specific'' enough to be considered, roughly speaking, the {\em
  loci} of competition between firms: the products belonging to a
given sub-market posses similar therapeutic characteristics and can
then be considered substitutable while products belonging to different
sub-markets are usually targeted to different pathologies. Moreover,
the data on licensing agreements~\citet{OPR} show that this disaggregation
level is also the one at which single research projects develop. These
characteristics lead to the conclusion that the different
``therapeutic sub-markets'' provide the natural ``scale'' at which the
``firm's growth'' phenomenon must be studied.

The analysis in this section is limited to the variance-size relation
characterizing the growth process of firms\footnote{For other aspect
  of the sales dynamics in sub-markets see \citet{BDLPR}}. It turns out
that the Law of Large Number is in fact responsible for this effect if
one assumes as ``atomic'' contribution to the firm's growth the growth
of firm's sales in the different sub-markets. This result comes from
two non-trivial observations: first, correlation across sub-market is
negligible and second, the number of active sub-markets of a given
firm is on average increasing with its size.

To make the argument clearer, let me introduce a bit of notation. Let
$S_{i,j}(t)$ be the size of firm $i$ in sub-market $j$
at time $t$. The aggregate size of the $i$-th firm is the sum of its
size on all the sub-market $S_{i}(t)=\sum_j S_{i,j}(t)$ and the
aggregate growth defined in \eqref{eq:gibrat} can be rewritten as
\begin{equation}
\label{eq:R_bysub}
R_i(t) + 1 = \frac{S_{i}(t+1)}{S_{i}(t)} = \sum_j \frac{S_{i,j}(t+1)}{S_{i}(t)} \,.
\end{equation}

If one computes the correlation of the ratios
$S_{(i,j)}(t+1)/S_{i}(t)$ for all the firms in all the sub-markets,
i.e. among all the possible couples of indeces $i$ and $j$, one
obtains a distribution centered around zero\footnote{With a standard
  deviation of $0.000388$ and an average deviation of $0.000024$}. The
growth processes on different sub-markets can be considered to any
extent uncorrelated and the variance of the aggregate growth is
obtained adding the variance of the growth in each sub-market.

\begin{figure}[tbp]
  \begin{center}
    \includegraphics[width=\myfigwidth]{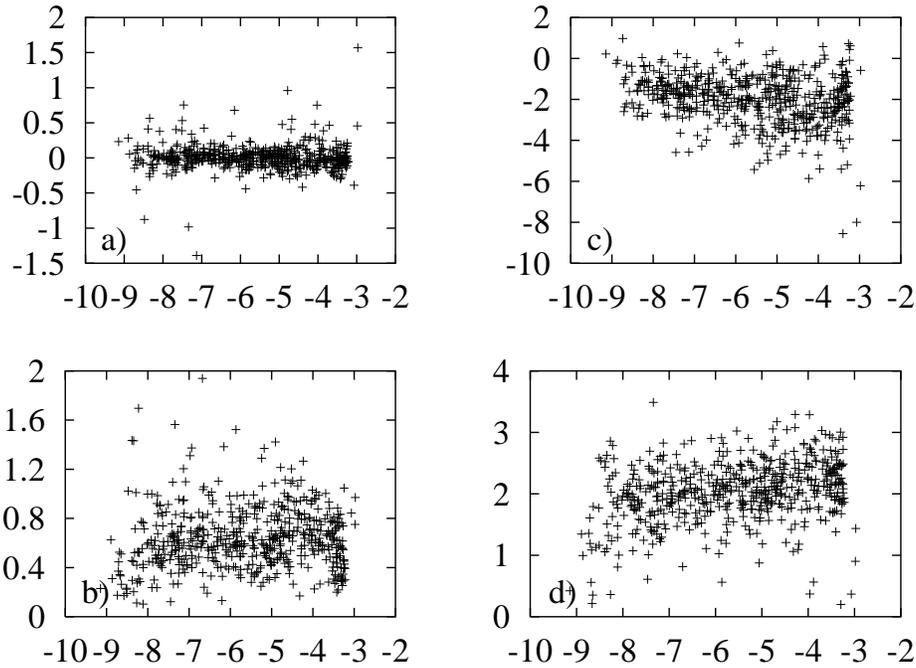}
    \caption{Results of various statistics computed on groups of firms binned
      according to their size. On the $x$-axis is the log of the
      average size of the bin. \textbf{(a)} Mean and \textbf{(b)}
      variance of $R_{i,j}(t)$ for different sub-markets $j$.
      \textbf{(c)} Mean and \textbf{(d)} variance of $\Delta_{i,j}(t)$
      for different sub-markets $j$.}
    \label{fig:portfolio}
  \end{center}
\end{figure}

In order to keep the present analysis consistent with the results of
previous Sections it is necessary to switch to the ``rescaled size''
$G_i(t)$.  This can be done straightforwardly but for clarity purposes
let me introduce the new variables
$R_{i,j}(t)=S_{i,j}(t+1)/S_{i,j}(t)$ and
$\Delta_{i,j}(t)=N_i(t)\,S_{i,j}(t)/S_{i}(t)$ where $N_i(t)$ is the
number of sub-markets in which firm $i$ operates at time $t$ (active
sub-markets). Then using \eqref{eq:R_bysub} the variance of the
``normalized growth'' can be written as
\begin{equation}
  \label{eq:G_growth}
  \mbox{var}_{i,t}[H_i(t)] = 
  \mbox{var}_{i,t} \left[ \frac{G_i(t+1)}{G_i(t)} \right]
  = \sum_j \mbox{var}_{i,t}\left[ \frac{M(t)}{M(t+1)} R_{i,j}(t)
    \frac{\Delta_{i,j}(t)}{N_i(t)}\right]
\end{equation}
where $M(t)$ is the average (aggregate) size of firms at time $t$ and
the ratio $M(t)/M(t+1)$ is a normalization factor (proportional to the
rate of growth of the total industry). I used the short notation
$\mbox{var}_{i,t}$ to denote the variance of the distribution
obtained, consistently to what done in previous Sections, using the
complete panel (all the firms at all the time steps).

\begin{figure}[tbp]
  \begin{center}
\setlength{\unitlength}{0.240900pt}
\begin{picture}(1500,900)(0,0)
\footnotesize
\thicklines \path(132,90)(152,90)
\thicklines \path(1433,90)(1413,90)
\put(110,90){\makebox(0,0)[r]{2.5}}
\thicklines \path(132,243)(152,243)
\thicklines \path(1433,243)(1413,243)
\put(110,243){\makebox(0,0)[r]{3}}
\thicklines \path(132,396)(152,396)
\thicklines \path(1433,396)(1413,396)
\put(110,396){\makebox(0,0)[r]{3.5}}
\thicklines \path(132,550)(152,550)
\thicklines \path(1433,550)(1413,550)
\put(110,550){\makebox(0,0)[r]{4}}
\thicklines \path(132,703)(152,703)
\thicklines \path(1433,703)(1413,703)
\put(110,703){\makebox(0,0)[r]{4.5}}
\thicklines \path(132,856)(152,856)
\thicklines \path(1433,856)(1413,856)
\put(110,856){\makebox(0,0)[r]{5}}
\thicklines \path(132,90)(132,110)
\thicklines \path(132,856)(132,836)
\put(132,45){\makebox(0,0){-9}}
\thicklines \path(318,90)(318,110)
\thicklines \path(318,856)(318,836)
\put(318,45){\makebox(0,0){-8}}
\thicklines \path(504,90)(504,110)
\thicklines \path(504,856)(504,836)
\put(504,45){\makebox(0,0){-7}}
\thicklines \path(690,90)(690,110)
\thicklines \path(690,856)(690,836)
\put(690,45){\makebox(0,0){-6}}
\thicklines \path(875,90)(875,110)
\thicklines \path(875,856)(875,836)
\put(875,45){\makebox(0,0){-5}}
\thicklines \path(1061,90)(1061,110)
\thicklines \path(1061,856)(1061,836)
\put(1061,45){\makebox(0,0){-4}}
\thicklines \path(1247,90)(1247,110)
\thicklines \path(1247,856)(1247,836)
\put(1247,45){\makebox(0,0){-3}}
\thicklines \path(1433,90)(1433,110)
\thicklines \path(1433,856)(1433,836)
\put(1433,45){\makebox(0,0){-2}}
\thicklines \path(132,90)(1433,90)(1433,856)(132,856)(132,90)
\put(1268,793){\raisebox{-1.2pt}{\makebox(0,0){$\Diamond$}}}
\put(1211,733){\raisebox{-1.2pt}{\makebox(0,0){$\Diamond$}}}
\put(1146,782){\raisebox{-1.2pt}{\makebox(0,0){$\Diamond$}}}
\put(1073,658){\raisebox{-1.2pt}{\makebox(0,0){$\Diamond$}}}
\put(983,704){\raisebox{-1.2pt}{\makebox(0,0){$\Diamond$}}}
\put(894,631){\raisebox{-1.2pt}{\makebox(0,0){$\Diamond$}}}
\put(834,512){\raisebox{-1.2pt}{\makebox(0,0){$\Diamond$}}}
\put(790,508){\raisebox{-1.2pt}{\makebox(0,0){$\Diamond$}}}
\put(753,526){\raisebox{-1.2pt}{\makebox(0,0){$\Diamond$}}}
\put(722,503){\raisebox{-1.2pt}{\makebox(0,0){$\Diamond$}}}
\put(683,511){\raisebox{-1.2pt}{\makebox(0,0){$\Diamond$}}}
\put(641,444){\raisebox{-1.2pt}{\makebox(0,0){$\Diamond$}}}
\put(613,449){\raisebox{-1.2pt}{\makebox(0,0){$\Diamond$}}}
\put(588,425){\raisebox{-1.2pt}{\makebox(0,0){$\Diamond$}}}
\put(569,361){\raisebox{-1.2pt}{\makebox(0,0){$\Diamond$}}}
\put(545,361){\raisebox{-1.2pt}{\makebox(0,0){$\Diamond$}}}
\put(520,282){\raisebox{-1.2pt}{\makebox(0,0){$\Diamond$}}}
\put(488,309){\raisebox{-1.2pt}{\makebox(0,0){$\Diamond$}}}
\put(459,362){\raisebox{-1.2pt}{\makebox(0,0){$\Diamond$}}}
\put(440,369){\raisebox{-1.2pt}{\makebox(0,0){$\Diamond$}}}
\put(424,239){\raisebox{-1.2pt}{\makebox(0,0){$\Diamond$}}}
\put(404,220){\raisebox{-1.2pt}{\makebox(0,0){$\Diamond$}}}
\put(386,218){\raisebox{-1.2pt}{\makebox(0,0){$\Diamond$}}}
\put(369,180){\raisebox{-1.2pt}{\makebox(0,0){$\Diamond$}}}
\put(349,144){\raisebox{-1.2pt}{\makebox(0,0){$\Diamond$}}}
\put(329,194){\raisebox{-1.2pt}{\makebox(0,0){$\Diamond$}}}
\put(310,195){\raisebox{-1.2pt}{\makebox(0,0){$\Diamond$}}}
\put(283,187){\raisebox{-1.2pt}{\makebox(0,0){$\Diamond$}}}
\put(249,205){\raisebox{-1.2pt}{\makebox(0,0){$\Diamond$}}}
\put(186,205){\raisebox{-1.2pt}{\makebox(0,0){$\Diamond$}}}
\thicklines \path(186,129)(186,129)(197,136)(208,143)(219,150)(230,157)(241,164)(252,171)(263,178)(273,185)(284,192)(295,199)(306,206)(317,213)(328,220)(339,228)(350,235)(361,242)(372,249)(383,256)(394,263)(405,270)(416,277)(427,284)(437,291)(448,298)(459,305)(470,312)(481,319)(492,326)(503,333)(514,340)(525,348)(536,355)(547,362)(558,369)(569,376)(580,383)(590,390)(601,397)(612,404)(623,411)(634,418)(645,425)(656,432)(667,439)(678,446)(689,453)(700,460)(711,467)(722,475)
\thicklines \path(722,475)(733,482)(743,489)(754,496)(765,503)(776,510)(787,517)(798,524)(809,531)(820,538)(831,545)(842,552)(853,559)(864,566)(875,573)(886,580)(896,587)(907,595)(918,602)(929,609)(940,616)(951,623)(962,630)(973,637)(984,644)(995,651)(1006,658)(1017,665)(1028,672)(1039,679)(1049,686)(1060,693)(1071,700)(1082,707)(1093,715)(1104,722)(1115,729)(1126,736)(1137,743)(1148,750)(1159,757)(1170,764)(1181,771)(1192,778)(1203,785)(1213,792)(1224,799)(1235,806)(1246,813)(1257,820)(1268,827)
\end{picture}
    \caption{Average number of active sub-markets. The fit is a
      straight line with slope $\alpha=.39+-.02$ and intercept $q=6.0
      +-0.12$}
    \label{fig:numerosity}
  \end{center}
\end{figure}
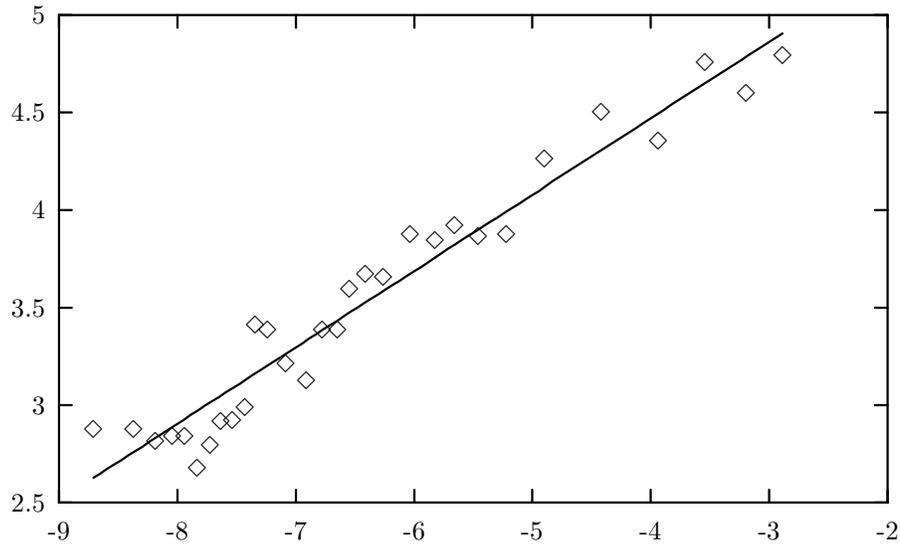

In \eqref{eq:G_growth} the contribution of each sub-market factorizes
in three terms: $R_{i,j}(t)$, which is the actual growth of the firm
$i$ in sub-market $j$; the inverse number of active markets $1/N_i(t)$
and $\Delta_{i,j}(t)$, which is a weighting coefficient describing the
``diversification'' heterogeneity of a firm: if the firm $i$ at time
$t$ is symmetrically diversified over its sub-markets, the
$\Delta_{i,j}(t)$ (for different $j$) are concentrated around $1$,
otherwise if the firm is asymmetrically diversified, the
$\Delta_{i,j}(t)$ are broadly distributed. The mean and variance of
the distributions for $R_{i,j}(t)$ and $\Delta_{i,j}(t)$ obtained
using different bins in the aggregate size do not show any clear
dependence on the average size of the bin (see
Fig.~\ref{fig:portfolio}). Then the term $N_i(t)$, the number of
active sub-market a firm posses, must be the sole responsible for the
observed dependence of variance over the aggregate size. Indeed,
fitting on a log-log scale the average number $<N_i(t)>$ of active
markets for each bin against the average size of the bin (see
Fig.~\ref{fig:numerosity}) one obtains a slope $\alpha = .39 +- .02$
and an intercept $q= 6 +- 0.12$.  The Law of Large Number would
predict a relation between the exponent in~\eqref{eq:varsize} and the
slope in Fig.~\ref{fig:numerosity} of the form $\beta = -\alpha/2$
which is in perfect agreement with the data\footnote{Notice that a
  weak relation appears between the variance of $\Delta_{i,j}$ and the
  total size. A linear fit provide a slope $.09$ that is negligible if
  compared to the effect due to the number of active sub-markets.}.

The conclusion is that the relation provided by the Law of Large
Number is valid as long as one consider the actual number of
sub-markets a firm operate in. It must be stress, however, that in
order to demonstrate this statement, it has been necessary to rule out
two possible sources of functional dependence between a firm's size
and the variance of its aggregate growth: the possibility that the
variance or the mean of a firm growth in a given sub-market depend
(on average) on its total size and the possibility that the
diversification patter of a firm (described by the variable
$\Delta_i$) varies (on average) with its size.  Both these
possibilities are actually discussed in literature \citet{HP} and
proposed as possible source of Gibrat Law violation.

It remains however to explain why the number of active markets a firm
possesses, shows a so clear dependence on its size $N(G) \sim
G^{\lambda}$ and what is the meaning (if any) of the parameter
$\lambda$. In the next Section I propose a model for the
diversification process of business firms, essentially based on the
assumption of a ``self-reinforcing'' effect concerning the
``proliferation'' of sub-markets a firm operates in, which provide a
simple but suggestive interpretation of Fig.~\ref{fig:numerosity}.

\section{A stochastic branching model for firm diversification}

\label{sec:branching}

The previous analysis shows that larger firms are present, on average,
on a greater number of sub-markets. For the purposes of this Section
it is useful to read this statement in a ``dynamical'' way: as firms
grow their activities becomes more and more diversified as they sell
products on different sub-markets.

In what follows I will try to capture the diversification
``behavior'' of firms with a model describing how the number of
active markets (or better, the probability of having a given number of
active markets) changes as firm grows. Notice that I'm not interested
in a ``dynamical'' process in time, i.e. a model describing the
evolution of diversification structure ``as time goes by''. This is
because the growth dynamics of the same firm could be very
heterogenous if observed at different time steps (actually, in this
sector, it is, see.\citet{BDLPR}) and it would be difficult to maintain
that shrinking firms behave, with respect to diversification, as
growing ones.  Rather, I will treat some ``size'' variable as
independent and describe the change in diversification patterns as
this quantity is varied. Due to the multiplicative nature of the
growth, the natural candidate to play the usual role of ``time'' is
the log of firms size $g$.  The object of the following analysis thus
becomes the probability that a firm which possesses $n$ active markets
when its (log) size is $g$, will possess $m$ active markets when its
(log) size is $g^\prime \geq g$, $P(n,g;m,g^\prime)$.

\begin{figure}[tbp]
  \begin{minipage}[t]{0.47\linewidth}
    \begin{center}
      \includegraphics[width=\textwidth]{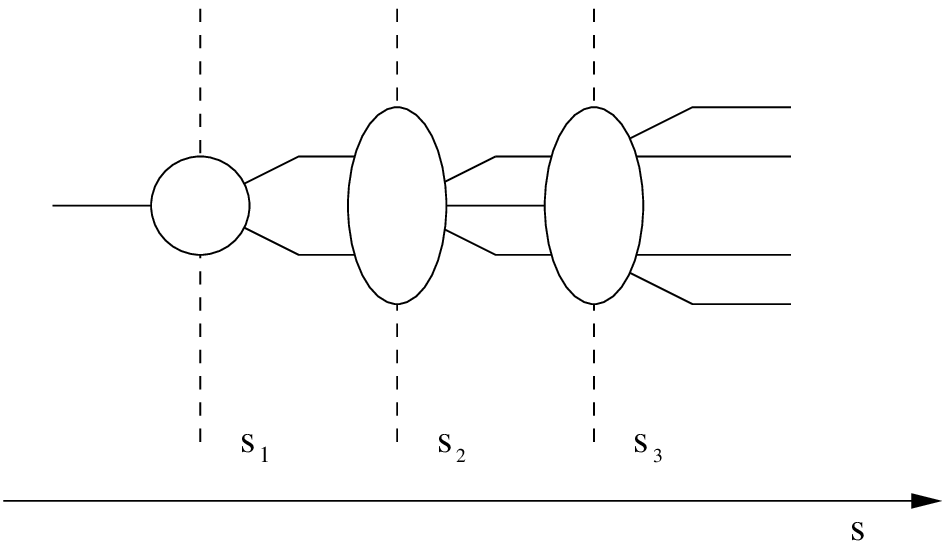}
      \caption{Diversification by independent events.
        The interarrival ``times'' $s_i-s_{i_1}$ are exponentially
        distributed.}
      \label{fig:poisson}
    \end{center}
  \end{minipage}
  \hfill
  \begin{minipage}[t]{0.47\linewidth}
    \begin{center}
      \includegraphics[width=\textwidth]{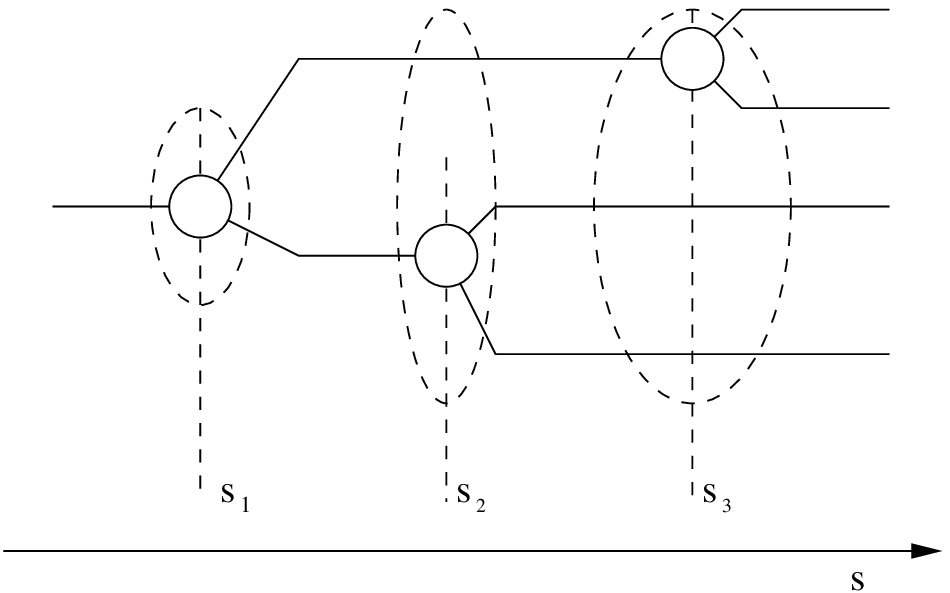}
      \caption{Diversification by branching. The interarrival
        ``times'' are no longer exponential, the distance between two
        events decreases as the number of active markets increases.}
      \label{fig:yule}
    \end{center}
  \end{minipage}
\end{figure}

Before proceeding with the model building, let me briefly mention a
collection of models that can be considered a previous attempt to
merge both the ``diversification'' and the ``growth'' aspect of firm's
dynamics. These models, collectively referred as ``island models'',
was originally proposed by Simon (see \citet{IS}), and can be
considered a ``classic'' in the stochastic growth theory of
firms\footnote{For their generality and robustness they have attracted
  attention also recently, for instance \citet{sutton2} is mainly based
  on this kind of models.}  There the ``diversification'' dynamics is
not directly analyzed, but rather serves as a ``driving'' process that
``sets the pace'' for the firms growth process, described as a
successive capture of diverse ``islands'', or ``business
opportunities''. In this way the distinction between ``growth'' and
``diversification'' becomes rather vague, but nevertheless two
assumptions seem to be generally accepted concerning the latter. The
first is on the nature of the ``diversification'' events (here the
entry on either a previously unexplored or unexploited market):
according to it, these events are seen as ``shocks'' (investment
opportunities, technological achievements, etc.) that can ``happen''
to firms along their histories with mutual independence. The second is
to consider these ``shocks'' uncorrelated in time. i.e.  to assume
that they could happen with the same constant probability in any
instant, irrespectively to the actual firm ``state''.  If one neglects
the possibility of ``instantaneous'' multiple
shocks\footnote{Equivalently: requires the property of orderliness for
  the associated stochastic process. For a general discussion on the
  complete characterization of point processes see \citet{snyder}}, the
previous assumptions provide a complete definition of the transition
probability $P(n,g;m,g+\delta)$ between two sizes differing by an
infinitesimal quantity $\delta$:
\begin{equation}
  \label{eq:poisson}
  P(n,g;m,g+\delta)=
  \left\{
    \begin{array}{ll}
      \lambda\,\delta + o(\delta)& m=n+1\\
      o(\delta)& m>n+1
    \end{array}
  \right. \;\;\;.
\end{equation}
where $\lambda$ is the constant ``shock rate'' (the average number of
``diversification events'' for unit time); this form of the transition
probability defines the well known Poisson process.  A pictorial
illustration of this process is shown in Fig.~\ref{fig:poisson}: the
``blobs'' stand for the diversification event ``black boxes'' that a
firm meets along its growth. As far as this model is concerned, it is
irrelevant what really happens inside the back box, what matters is
that after the ``shock'' the firm ends with one more active
sub-market.

\begin{figure}[tbp]
  \begin{center}
    \includegraphics[width=\myfigwidth]{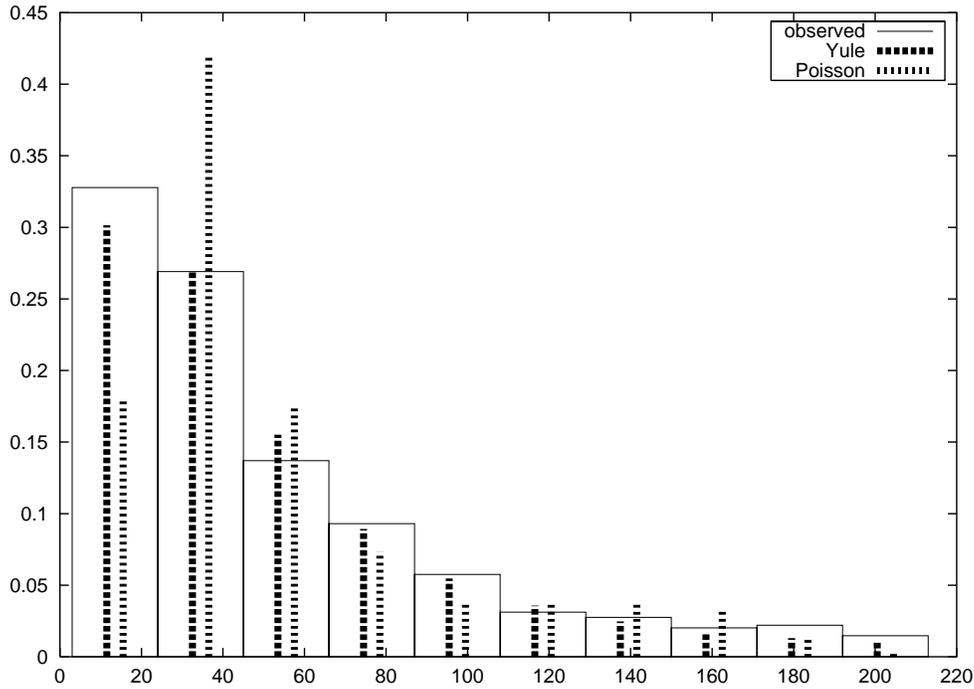}
    \caption{The binned probability density for the number of sub-market
      computed directly by the data and theoretically predicted by the
      Yule process. The fit showed was obtained with $\lambda=.39$,
      $n_0=5$ and $s_0=-12$. As comparison, a fit of a Poisson model
      with a non linear intensity function $\tau(G)=G^\lambda$ is also
      shown.}
    \label{fig:yule_test}
  \end{center}
\end{figure}

The Poisson process would however predict a linear increase of the
average number of active sub-markets with $g$, a property that is
clearly in contrast with the empirical results discussed in the
previous Sections\footnote{Notice that one could also account for the
  observed behavior using a Poisson process with a non linear
  intensity function $\tau(G)=G^\lambda$ (see \citet{snyder}). This
  approach seems however rather unnatural and would provide a worse
  description for the overall statistics, see
  Fig.\ref{fig:yule_test}}. In order to obtain a more satisfying
model, it proves to be useful to explore inside the ``black boxes''
of Fig~\ref{fig:poisson}, and try to describe, at least partially, the
actual nature of the diversification events.  A possible
interpretation of these events comes from the knowledge about the
``technological'' behavior of the industry~\citet{OPR}. The growth in
pharmaceutical industry is highly ``research driven''\footnote{Roughly
  speaking, this means that a relative large amount of firms budget is
  invested in formal R\&D activities and these activities generate,
  via products innovations, increasing competitiveness for the firms}
so that it is plausible to think to the diversification too as an
outcome of ``technical improvements'': a firm enters a new market when
it has the (technical) capability of developing products for this
market.  Moreover previous studies over a wide set of patent
agreements \citet{OPR} showed that, for what concerns this sector, the
technical knowledge proceeds with successive refinements with new
results bringing in new possibilities of technological advancements.
Then, the simplest way of mapping the resulting ``technological
behavior'' on the diversification dynamics is to suppose that the
diversification process proceed as a branching process, where each
opened branch (sub-market) becomes eventually the origin of a new
branching (diversification) event. A picture of this process is shown
in Fig.~\ref{fig:yule} under the (minimal) assumption that the
branching is uniformly binary. If one neglects the topology in
sub-markets space that emerges from these successive branchings, but
is interested only in the actual number of these sub-markets, the
process in Fig.~\ref{fig:yule} can be readily described: all the
active sub-markets can be sources of a possible ``diversification''
event ``\`a la'' Poison, and the~\eqref{eq:poisson} must be modified
to read
\begin{equation}
  \label{eq:yule}
  P(n,g;m,g+\delta)=
  \left\{
    \begin{array}{ll}
      n\,\lambda\,\delta + o(\delta)&m=n+1\\
      o(\delta) &m>n+1
    \end{array}
  \right.
\end{equation}
where again multiple instantaneous branchings are neglected. If
$p_n(g)$ is the probability that a firm of size $g$ has $n$ active
markets it must satisfy the (pure-birth) set of equations
\begin{equation}
  \label{eq:master}
  \begin{array}{lclr}
    p_n(g+\delta) &=&
    p_{n-1}(g)\,P(n-1,g;n,g+\delta)-p_n(g)\,P(n,g;n,g+\delta) & n>n_0 \\
    p_{n_0}(g+\delta) &=& -p_{n_0}(g)\,P(n_0,g;n_0,g+\delta)& n=n_0
  \end{array}
\end{equation}
where $n_0$ is the initial number of active markets. Substituting in
\eqref{eq:master} the definition given in \eqref{eq:yule} and taking
the limit $\delta \to 0$ one obtains the set of differential
equations:
\begin{equation}
  \label{eq:yule_ode}
  \begin{array}{lclr}
    p_n^\prime(g)&=&-n\,\lambda\,p_{n}(g)+(n-1)\,\lambda\, p_{n-1}(g)&n>n_0\\
    p_{n_0}^\prime(g) &=& -n_0\,\lambda\,p_{n_0}(g)& n=n_0
  \end{array}
\end{equation}
with the initial conditions:
\begin{equation}
  \label{eq:yule_ode_inicond}
  p_n(g_0) = \left\{
    \begin{array}{l}
      1, \; n=n_0\\
      0, \; n \neq n_0
    \end{array}
  \right.
\end{equation}
where $g_0$ is the initial size of the firm. The previous process is
known as the Yule process and has been originally proposed to explain
the proliferation in time of animal species (for a discussion see
\citet{Feller} and reference therein). The system \eqref{eq:yule_ode} can
be easily solved (see Appendix 1) to obtain the following
distribution:
\begin{equation}
  \label{eq:yule_distrib}
  p_n(g) = {n-1 \choose n-n_0} \, e^{-n_0 \lambda (g-g_0)} \, \left(
    1-e^{-\lambda (g-g_0)} \right)^{n-n_0} \;,
\end{equation}
defined for $g \geq g_0$ and $n \geq n_0$.

Let me turn now to the problem of describing the observed data with
the proposed model. The distribution in \eqref{eq:yule_distrib}
contains three parameters: the ``diversification'' rate $\lambda$, the
initial number of active sub-markets $n_0$ and the initial firm size
$s_0$. Since it correctly predicts an exponential increase in the
average number of active markets with size
\begin{equation}
  \label{eq:yule_mean}
  <N(g)> = \sum_{n=n_0} n p_n (g) = n_0 e^{\lambda (g-g_0)} \;,
\end{equation}
one can use the linear fit shown in Fig.~\ref{fig:numerosity} to obtain
an estimate of the parameters in \eqref{eq:yule_mean}, requiring the
fulfillment of the two conditions $\lambda=\alpha=.39$ and $\log(n_0)
- \lambda s_0=q=-6$.

The descriptive power of the model, however, must be judge using 
more distinguishing object than the average $<N(s)>$. A good candidate
is the diversification pattern of the whole industry. To be precise,
let $P(M_0<M<M_1)$ be the probability that a firm posses a number of
active market between $M_0$ and $M_1$.  This quantity can be evaluated
using the actual frequency of occurrences:
\begin{equation}
  \label{eq:prob_number}
  P^{ob}(M_0<M<M_1) = \frac{1}{T N} \sum_{m=M_0}^{M_1} \sum_{i=1}^{N} 
  \sum_{t=0}^{T-1} \delta(n_i(t) - m)
\end{equation}
where $n_i(t)$ is the number of active market of firm $i$ at time $t$
and $\delta$ is the Kronecker delta. There is, however, an alternative
approach for the computation of the previous quantity: it can be
obtained via a simple convolution
\begin{equation}
  \label{eq:covolution}
  P(M_0<M<M_1) = \sum_{m=M_0}^{M_1} \int_{g_0}^{+\infty} dg
  f(g)\,p_m(g)
\end{equation}
once the size probability density $f(s)$ and the ``diversification''
probability $p_n(g)$ are known. An estimate of \eqref{eq:covolution}
can be obtained using the observed frequencies in data
\begin{equation}
  \label{eq:prob_number_yule}
  P^{th}(M_0<M<M_1) = \frac{1}{T N} \sum_{m=M_0}^{M_1} \sum_{i=1}^{N}
  \sum_{t=0}^{T-1} p_m (g_i(t))
\end{equation}
where $g_i(t)$ is the size of firm $i$ at time $t$. Notice that this
second method depends on the ``branching'' model to obtain an estimate
of the probabilities $P(M_0<M<M_1)$. A comparison between the observed
$P^{ob}$ and the predicted $P^{th}$ distributions provides a way to
tune the residual degree of freedom, and an indication of the
goodness of the model. As showed in Fig.~\ref{fig:yule_test} the
accordance, considering the extreme simplifications introduced, is
good.

\section{Conclusions and outlook}

The relation between the variance of growth and the size, which
constitutes the clearest and often reported violation of the Gibrat
law has been reduced to a diversification effect: bigger firms operate
in more sub-markets and the variance of their growth is consequently
reduced\footnote{Incidentally, the observed relation between the
  number of sub-markets a firm posses and its size constitutes an
  original example of a ``scaling'' relation, and has to be added to
  other more famous example pertaining the domain of economics (for a
  review and critical discussion see \citet{Brock2})}. I have shown
that, in the industry under analysis, other possible effects (in
particular the correlation among sub-markets and the dependence of the
``strategic'' diversification pattern of a firm on its size) if exist,
are negligible. 

The actual structure of the firm diversification can be described with
good accuracy using the simple stochastic process proposed in
Sec.~\ref{sec:branching}.  A {\em caveat} must be introduced, however,
regarding the ``interpretation'' of this model: the strong
technological component in industry growth suggests, but does not
imply, that the main driving force in diversification would be some
sort of ``technological specification''. The ``penetration'' of a firm
in a new sub-market can actually possess various natures, nevertheless
the proposed model is likely to describe the diversification dynamics
irrespectively of this variety. The reasons of this flexibility are in
the simplicity of the hypotheses. The model is indeed based on two
general assumptions: first, the existence of some sort of
``competencies'' providing the firm with the ability to diversify its
business, and second, that these ``competencies'' (and so the ability
to enter a new sub-market) increase with the number of times they have
been effectively used (and so with the number of opened
sub-markets)\footnote{Note that in the model the cumulative effect on
  ``competencies'' is described by a linear function. In general it
  would be probably possible to obtain a better agreement using more
  sophisticated relations and processes, for instance dropping the
  strictly binary nature of the branching or introducing the
  possibility of a branch ``death''. This ``finer'' modeling would
  require, however, a higher ``phenomenological'' justification from
  the data.}.  Both these assumption seem, in most cases, so
reasonable that would be difficult to find arguments against them.


The previous model, thus, suggests an ``evolutionary'' description of
firms which progressively ``learn'' to diversify, whatever the object
of this learning would be. Incidentally, the observed relationship
between the growth variance and the size of the firms, being milder
than the LLN prediction, gives evidence against the interpretation of
``diversification'' as a risk minimization strategy: indeed if this
would be the case, firms have to be present on much more sub-markets
then they actually are.

The present analysis can be extended in several directions: the first
concerns the generality of the previous findings relating Gibrat's Law
violation to a diversification effect. In Sec.~\ref{sec:gibrat} only
one industry has been analyzed, but the observed trend in
Fig.~\ref{fig:bysize} is so similar to other results in literature
(\citet{stanley1},\citet{stanley2}) that a major multi-sectorial
investigation, where possible, is advisable. Second, concerning the
proposed model of firm diversification, it would be important to
empirically investigate the nature of the ``competencies'' leading to
penetration in new markets and their connection with the different
characteristics of the firm, the more interesting probably being its
``technological advantage''.  This investigation would constitute an
important empirical support to the construction of evolutionary models
of the firm (in line with what has been done, for instance, in
\citet{DMOS}).  Finally, it would be interesting to apply the same
description to diversification data from other industries.  This would
be a test of the model previously described, but, more interestingly,
could constitute, via the introduction of industry-specific
parameters, a ``dynamical'' way of characterizing diversification
pattern in different sectors.

\section{Acknowledgments}
The author thanks Giovanni Dosi, Fabio Pamolli and Massimo Riccaboni
for helpful comments and useful discussions.

\newpage

\begin{center}
  {\bf APPENDIX}
\end{center}

\appendix

\section{Solution of the Yule process}

To simplify the solution of \eqref{eq:yule_ode} let me introduce a
rescaled size $t=\lambda(g-g_0)$ and consider, instead of the
probabilities in \eqref{eq:yule}, the variables $y_n(t)$ defined by
\begin{equation}
  \label{eq:factor_p}
  p_n(t) = e^{-n t} \, y_n(t) \;.
\end{equation}
In these new variables the set of equation in \eqref{eq:yule_ode}
reduces to:
\begin{equation}
  \label{eq:g_ode}
  \begin{array}{lclr}
    y^\prime_n (t) &=& (n-1)\, e^{t}\, y_{n-1}(t) & n>n_0\\
    y^\prime_{n_0}(t) &=& 0 & n=n_0 \;.
  \end{array}  
\end{equation}
with initial conditions:
\begin{equation}
  \label{eq:g_ode_inicond}
  y_n(0) = \left\{
    \begin{array}{l}
      1, \; n=n_0\\
      0, \; n \neq n_0
    \end{array}
  \right.
\end{equation}
From \eqref{eq:g_ode}, iterating over the index $n$, it is immediate
to write the solution as multiple integral:
\begin{equation}
  \label{eq:itersoll}
  y_n(t) = (n-1)\,(n-2)\ldots n_0 \,
  \int_0^{t} dt_1 \int_0^{t1} dt_2 \ldots 
  \int_0^{t_{n-n_0+1}} dt_{n-n_0}
  e^{t_1 + t_2 + \ldots + t_{n-n_0}}
\end{equation}
Notice that, due to the complete symmetry of the integrand, the
multiple integral over the $n-n_0$-dimensional hyper-cube of side $t$
with the constraints $t_{n-n_0} < \ldots < t_2 < t_1 < t$ reduces to
the integral over the whole hypercube divided by all the possible
ordering of the constraints, which are $(n-n_0)!$. One thus obtains:
\begin{equation}
  \label{eq:g_result}
  y_n(t) = {n-1 \choose n-n_0}\,(e^t-1)^{n-n_0} \;,
\end{equation}
that, remembering the factorization in \eqref{eq:factor_p} and
substituting $t$ with its previous definition, reduces
to~\eqref{eq:yule_distrib} after obvious algebra.

\end{document}